\documentclass[prb,twocolumn,groupedaddress,shownopacs,amssymb]{revtex4-2}
\usepackage{graphicx,psfrag,amsmath,amssymb}

\usepackage[usenames, dvipsnames]{color}

\usepackage{hyperref} 
\hypersetup{
    colorlinks=true,
    linkcolor=Blue,
    citecolor=Blue,
    filecolor=Blue,      
    urlcolor=Blue,
    pdftitle={structure factor},
    }

\begin{document}

\title{Structure factors and quantum geometry in multiband BCS superconductors}

\author{M. Iskin}
\affiliation{
Department of Physics, Ko\c{c} University, Rumelifeneri Yolu, 
34450 Sar\i yer, Istanbul, T\"urkiye
}

\date{\today}

\begin{abstract}

We consider multiband BCS superconductors that exhibit time-reversal 
symmetry and uniform pairing, and analyze their dynamic density and 
spin structure factors using linear-response theory within the 
mean-field BCS-BEC crossover framework at zero temperature. Our results 
for the multi-orbital Hubbard model satisfy the associated f-sum rules 
in several limits. In particular, in the strong-coupling limit, they 
coincide with those of a weakly-interacting Bose gas of Cooper pairs, 
where the low-energy collective Goldstone modes serve as 
Bogoliubov phonons. We further reveal that the 
quantum-geometric origin of the low-energy structure factors, along with 
related observables such as the superfluid-weight tensor and the 
effective-mass tensor of Cooper pairs, can be traced all the way back 
to the effective-mass theorem for Bloch bands in this limit. 
As an illustration, we investigate the pyrochlore-Hubbard model 
numerically and demonstrate that the Goldstone modes are the only relevant 
collective degrees of freedom in the flat-band regime. 

\end{abstract}
\maketitle

\section{Introduction}
\label{sec:intro}

Advances in solid-state and condensed-matter physics have increasingly 
demonstrated the importance of quantum geometry in understanding and 
controlling the behavior of multiband 
systems~\cite{tan19, yu20, gianfrate20, tian23, yi23, gao23, 
kim25, sala24, kang24, tanaka24}. 
Central to this perspective is the complex quantum-geometric 
tensor~\cite{Provost80, berry84, resta11}, whose imaginary 
part, i.e., the Berry-curvature tensor, has long been recognized for 
its role in defining topological invariants such as the Chern number. 
This invariant not only classifies topological phases of matter 
but also governs transport phenomena in materials ranging from 
topological insulators to unconventional superconductors.
Equally compelling is the real part of the quantum-geometric tensor, 
i.e., the quantum-metric tensor, which quantifies the quantum distance 
between nearby Bloch states. While historically less explored 
than its Berry-curvature counterpart, the quantum metric has 
emerged as a crucial factor in a variety of physical 
contexts~\cite{torma22, torma23, yu24, jiang25}. 
In multiband superconductors, for example, it mediates virtual 
interband processes that endow Cooper pairs with a finite 
effective mass even in systems where an isolated flat band 
would otherwise preclude superconductivity due to an infinite 
band mass. This geometric mechanism thus provides a natural 
explanation for the emergence of superconductivity in flat-band 
systems, linking key observables such as the superfluid weight, 
critical temperature, coherence length, pair size and low-energy 
collective modes. For an extended bibliography on this topic, see the
recent reviews~\cite{jiang25, torma22, torma23, yu24}.

The interplay between conventional band-dispersion effects and these 
geometric contributions becomes particularly pronounced in nearly 
flat bands, where the former terms are suppressed. 
Recent theoretical work, spanning from one-dimensional to three-dimensional 
tight-binding toy models has solidified the notion that nontrivial 
band geometry is indispensable for realizing superconductivity in 
flat-band materials~\cite{torma22, torma23, yu24, jiang25}. By integrating
both topological and geometric insights, this emerging
framework challenges conventional BCS paradigms and opens new avenues
for designing high-critical-temperature superconductors. Ultimately, 
the synthesis of quantum-geometric concepts with traditional band theory 
offers a unified perspective on superconducting pairing, illuminating 
the subtle yet profound influence of band geometry on the 
macroscopic properties of quantum materials. 
Thus, motivated by recent studies on the static structure factor of 
band insulators~\cite{onishi24b, onishi24c}, here we utilize 
linear-response theory within the mean-field BCS-BEC crossover framework 
at zero temperature, and examine the dynamic density structure factor (DSF) 
and dynamic spin structure factor (SSF) of multiband BCS superconductors 
that exhibit time-reversal symmetry and uniform pairing. Our analysis of the 
multi-orbital Hubbard model reveals that the collective-mode contribution 
to the spectral function of the density response is directly related to 
the superfluid-weight tensor. In the strong-coupling limit, they align 
with those of a weakly interacting Bose gas of Cooper pairs, 
where the low-energy collective Goldstone modes serve as Bogoliubov phonons. 
Furthermore, we show that the quantum-geometric 
origins of the low-energy structure factors, superfluid-weight tensor 
and the effective-mass tensor of Cooper pairs can all be traced 
back to the effective-mass theorem for Bloch bands in this limit.

The remainder of the paper is organized as follows. 
In Sec.~\ref{sec:cf}, we introduce the density and spin response 
functions, along with their associated structure factors and 
f-sum rules, for a generic multi-orbital Hubbard model. 
In Sec.~\ref{sec:lrt}, we discuss the zero-temperature linear-response 
theory for multiband BCS superconductors and analyze the structure 
factors within the mean-field BCS-BEC crossover framework. 
In Sec.~\ref{sec:ni}, we present numerical calculations for the 
pyrochlore-Hubbard model. In Sec.~\ref{sec:ci}, we comment on the role
of Coulomb interactions on the low-lying collective excitations. 
The paper concludes with a summary in Sec.~\ref{sec:conc}, 
while a brief discussions of the effective Gaussian action, 
the Ward identities, and the effective-mass tensors are provided 
in the Appendix.

\section{Correlation Functions}
\label{sec:cf}

The linear density-response function 
$
\mathcal{X}(\mathbf{r}, t) = -i\theta(t) 
\langle [\rho(\mathbf{r}, t), \rho(\mathbf{0}, 0)] \rangle
$
measures the correlation between the density of particles at two different 
points, where $\theta(t)$ is the Heaviside step function, $\langle \cdots \rangle$ 
represents the thermal statistical average, and
$
\rho(\mathbf{r}, t) = \rho_\uparrow(\mathbf{r}, t) + \rho_\downarrow(\mathbf{r}, t)
$
is the total density operator in the Heisenberg picture~\cite{nozieres1999theory}. 
Its spatial Fourier transform is given by
$
\mathcal{X}(\mathbf{q}, t) = \frac{1}{\sqrt{V}} \int d^3\mathbf{r} 
e^{-i \mathbf{q} \cdot \mathbf{r}} \mathcal{X}(\mathbf{r}, t),
$
which can be obtained using
$
\rho(\mathbf{r}, t) = \frac{1}{\sqrt{V}} \sum_\mathbf{q} 
e^{i \mathbf{q} \cdot \mathbf{r}} \rho(\mathbf{q}, t)
$
along with
$
\int d^3\mathbf{r} e^{i(\mathbf{q-q'})\cdot \mathbf{r}} 
= V \delta_{\mathbf{q}\mathbf{q'}}.
$
Here, $V$ is the volume of the system and $\delta_{ij}$ is a Kronecker delta.
The spectral function of its spatio-temporal Fourier transform,
\begin{align}
\mathcal{X}(\mathbf{q}, \omega) = -i \int_0^\infty dt e^{i \omega t}
\langle [\rho(\mathbf{q}, t), \rho(\mathbf{-q}, 0)] \rangle,
\label{eqn:defX}
\end{align}
determines the dynamic and static DSFs as follows.

\subsection{Density structure factor (DSF)}
\label{sec:dsf}

The imaginary part of the density response function is related to the 
dynamic DSF $S(\mathbf{q}, \omega)$ through
$
\mathrm{Im} \mathcal{X}(\mathbf{q}, \omega + i0^+) = - \pi 
[S(\mathbf{q}, \omega) - S(-\mathbf{q}, -\omega)],
$
where
$
S(\mathbf{q}, -\omega) = e^{-\beta \omega} S(\mathbf{q}, \omega)
$
is the so-called detailed balance~\cite{nozieres1999theory}. 
Throughout this paper, we assume inversion symmetry and set
$
S(-\mathbf{q}, \omega) = S(\mathbf{q}, \omega),
$
which leads to
$
S(\mathbf{q}, \omega) = - \frac{1}{\pi} \frac{\mathrm{Im} 
\mathcal{X}(\mathbf{q}, \omega + i0^+)}{1 - e^{-\beta \omega}},
$
where $\beta = 1/T$ is the inverse temperature in units of $\hbar = k_B = 1$ 
the Planck and Boltzmann constants. Furthermore, at zero temperature 
(of interest in this paper), one finds~\cite{nozieres1999theory}
\begin{align}
S(\mathbf{q}, \omega) = - \frac{1}{\pi} \mathrm{Im} 
\mathcal{X}(\mathbf{q}, \omega + i0^+)
\label{eqn:defS}
\end{align}
for $\omega > 0$, and $S(\mathbf{q}, \omega) = 0$ for $\omega < 0$.
By plugging 
$
\rho(\mathbf{q}, t) = e^{i \mathcal{H} t} \rho_\mathbf{q} e^{-i \mathcal{H} t}
$
into Eq.~(\ref{eqn:defX}), where
$
\rho_\mathbf{q} = \rho_{\uparrow \mathbf{q}} + \rho_{\downarrow \mathbf{q}}
$
is the Fourier transform of the total density operator at $t = 0$, 
and $\mathcal{H}$ is the Hamiltonian that governs the system, 
and using the Schr\"odinger equation
$
\mathcal{H} |\phi_\mathcal{N} \rangle = \mathcal{E}_\mathcal{N} |\phi_\mathcal{N} \rangle
$
for the stationary states along with the completeness relation
$
\sum_\mathcal{N} |\phi_\mathcal{N} \rangle \langle \phi_\mathcal{N} | = 1,
$
we find
$
\mathrm{Im} \mathcal{X}(\mathbf{q}, \omega + i0^+) = - \pi \sum_\mathcal{N} 
[
|\langle \phi_0| \rho_\mathbf{q} | \phi_\mathcal{N} \rangle|^2
\delta(\omega - \mathcal{E}_\mathcal{N} + \mathcal{E}_0)
- |\langle \phi_0| \rho_\mathbf{q}^\dagger | \phi_\mathcal{N} \rangle|^2
\delta(\omega + \mathcal{E}_\mathcal{N} - \mathcal{E}_0)
]
$
at $T = 0$. Here, $|\phi_0 \rangle$ represents the ground state and we make 
use of the identity
$
1/(x + i0^+) = \mathcal{P}(\frac{1}{x}) - i\pi \delta(x),
$
with $\mathcal{P}$ denoting the Cauchy principal value and $\delta(x)$ the
Dirac delta function.
Thus, the dynamic DSF is simply given by
$
S(\mathbf{q}, \omega) = \sum_\mathcal{N} 
|\langle \phi_0| \rho_\mathbf{q} | \phi_\mathcal{N} \rangle|^2
\delta(\omega - \mathcal{E}_\mathcal{N} + \mathcal{E}_0),
$
leading to the so-called f-sum rule
$
\int_0^\infty d\omega \omega S(\mathbf{q},\omega) = 
\sum_\mathcal{N} (\mathcal{E}_\mathcal{N}-\mathcal{E}_0)
|\langle \phi_0| \rho_\mathbf{q} | \phi_\mathcal{N} \rangle|^2.
$
It is easy to show that
$
\langle \phi_0| [\rho_\mathbf{q}^\dagger, [\rho_\mathbf{q}, \mathcal{H}]] 
|\phi_0 \rangle = - \sum_\mathcal{N} (\mathcal{E}_\mathcal{N}-\mathcal{E}_0) 
[
|\langle \phi_0| \rho_\mathbf{q} |\phi_\mathcal{N} \rangle|^2
+ |\langle \phi_0| \rho_\mathbf{q}^\dagger |\phi_\mathcal{N} \rangle|^2
]
$
in general, where 
$
\rho_\mathbf{q}^\dagger = \rho_{-\mathbf{q}},
$
and this can be expressed as~\cite{nozieres1999theory} 
\begin{align}
\label{eqn:fsc}
\langle \phi_0| [\rho_{\mathbf{q}}^\dagger, 
[\rho_{\mathbf{q}}, \mathcal{H}]] |\phi_0 \rangle
= - 2\int_0^\infty d\omega \omega S(\mathbf{q},\omega)
\end{align}
in the presence of inversion symmetry. Furthermore, the static DSF
$S(\mathbf{q})$ is defined by
$
\int_0^\infty S(\mathbf{q}, \omega) d\omega = \langle \phi_0|
\rho_\mathbf{q} \rho_\mathbf{q}^\dagger|\phi_0 \rangle 
= \frac{n_0}{2} S(\mathbf{q}),
$
where $n_0$ is the total density of particles~\cite{combescot06, he16}.

\subsection{Spin structure factor (SSF)}
\label{sec:ssf}

Similar to Eq.~(\ref{eqn:defX}), we define the linear spin-response function 
for the spin-density operator
$
\rho_s(\mathbf{r}, t) = \rho_\uparrow(\mathbf{r}, t) - \rho_\downarrow(\mathbf{r}, t)
$
as
$
\mathcal{X}_s(\mathbf{q}, \omega) = -i \int_0^\infty dt e^{i \omega t}
\langle [\rho_s(\mathbf{q}, t), \rho_s(\mathbf{-q}, 0)] \rangle,
$
which leads to
$
S_s(\mathbf{q}, \omega) = - \frac{1}{\pi} \mathrm{Im} 
\mathcal{X}_s(\mathbf{q}, \omega + i0^+)
$
for $\omega > 0$, and $S_s(\mathbf{q}, \omega) = 0$ for $\omega < 0$
at $T = 0$. It is easy to show that 
$
[\rho_{s \mathbf{q}}^\dagger, [\rho_{s \mathbf{q}}, \mathcal{H}]] 
= [\rho_{\mathbf{q}}^\dagger, [\rho_{\mathbf{q}}, \mathcal{H}]] 
$
for the operator
$
\rho_{s \mathbf{q}} = \rho_{\uparrow \mathbf{q}} -  \rho_{\downarrow \mathbf{q}},
$
leading to the spin f-sum rule~\cite{guo13, he16}
\begin{align}
\label{eqn:fsrs}
\int_0^\infty d\omega \omega S_s(\mathbf{q},\omega)
= \int_0^\infty d\omega \omega S(\mathbf{q},\omega).
\end{align}
Alternatively, one can define the spin-resolved dynamic structure 
factors~\cite{combescot06, kitamura24}
$
S_{\sigma \sigma'}(\mathbf{q}, \omega) = \sum_\mathcal{N} 
\langle \phi_0| \rho_{\sigma \mathbf{q}} | \phi_\mathcal{N} \rangle
\langle \phi_\mathcal{N}| \rho_{\sigma' \mathbf{q}}^\dagger | \phi_0 \rangle
\delta(\omega - \mathcal{E}_\mathcal{N} + \mathcal{E}_0),
$
and identify 
$
S(\mathbf{q}, \omega) = \sum_{\sigma \sigma'} S_{\sigma \sigma'}(\mathbf{q}, \omega) 
$
as the dynamic DSF and
$
S_s(\mathbf{q}, \omega) = \sum_{\sigma \sigma'} (-1)^{1+\delta_{\sigma \sigma'}} 
S_{\sigma \sigma'}(\mathbf{q}, \omega)
$
as the dynamic SSF. It is expected that
$
S_{\uparrow \uparrow}(\mathbf{q}, \omega) 
= S_{\downarrow \downarrow}(\mathbf{q}, \omega) 
$
and
$
S_{\uparrow \downarrow}(\mathbf{q}, \omega) 
= S_{\downarrow \uparrow}(\mathbf{q}, \omega) 
$
in the presence of time-reversal and inversion symmetries, which 
are assumed to be the case in this paper. This suggests that
$
\int_0^\infty d\omega \omega S_{\uparrow \downarrow}(\mathbf{q},\omega) = 0
$
~\cite{he16}.
Furthermore, 
$
\langle \phi_0| [\rho_{\sigma \mathbf{q}}^\dagger, 
[\rho_{\sigma' \mathbf{q}}, \mathcal{H}]] |\phi_0 \rangle
= - \int_0^\infty d\omega \omega [S_{\sigma \sigma'}(-\mathbf{q},\omega)
+ S_{\sigma' \sigma}(\mathbf{q},\omega)]
$
in general, and the static SSF $S_s(\mathbf{q})$ is defined by
$
\int_0^\infty S_s(\mathbf{q}, \omega) d\omega = \langle \phi_0|
\rho_{s \mathbf{q}} \rho_{s \mathbf{q}}^\dagger|\phi_0 \rangle 
= \frac{n_0}{2} S_s(\mathbf{q})
$
~\cite{combescot06, he16}.

\subsection{F-sum rule in a multi-orbital Hubbard model}
\label{sec:tbl}

Next, we consider a multiband tight-binding lattice that is described by 
the Bloch Hamiltonian matrix $\mathbf{h}_{\mathbf{k} \sigma}$, which 
satisfies the Schr\"odinger equation
$
\mathbf{h}_{\mathbf{k} \sigma} | n_{\mathbf{k} \sigma} \rangle
= \varepsilon_{n \mathbf{k} \sigma} | n_{\mathbf{k} \sigma} \rangle
$
for the periodic part $| n_{\mathbf{k} \sigma} \rangle$ of the Bloch 
states. Here, the dispersion relation $\varepsilon_{n \mathbf{k} \sigma}$ 
for the $n$th Bloch band is determined by the eigenvalue relation
\begin{align}
\sum_{S'} h_{\mathbf{k} \sigma}^{SS'} n_{S' \mathbf{k} \sigma}
= \varepsilon_{n \mathbf{k} \sigma} n_{S \mathbf{k} \sigma},
\end{align}
where
$
n_{S \mathbf{k} \sigma} = \langle S| n_{\mathbf{k} \sigma} \rangle
$
is the projection to the $S$th sublattice in a unit cell.
The precise form of the matrix elements $h_{\mathbf{k} \sigma}^{SS'}$ 
depends on the lattice geometry and hopping parameters. 
Suppose the operator $c_{S \mathrm{i} \sigma}^\dagger$ creates a 
spin-$\sigma$ fermion in the $\mathrm{i}$th unit cell on sublattice 
$S \in \{A, B, C, \cdots \}$. 
Then, the corresponding site operator for the number of spin-$\sigma$ 
fermions is given by
$
\rho_{S \mathrm{i} \sigma} = c_{S \mathrm{i} \sigma}^\dagger c_{S \mathrm{i} \sigma}.
$
Using the discrete Fourier transforms
$
c_{S \mathrm{i} \sigma} = \frac{1}{\sqrt{N_c}} \sum_{\mathbf{k}} 
e^{i \mathbf{k} \cdot \mathbf{r}_{S \mathrm{i}}} c_{S \mathbf{k} \sigma}
$
and
$
\rho_{S \mathbf{q} \sigma} = \frac{1}{\sqrt{N_c}} \sum_{\mathrm{i}} 
e^{-i \mathbf{q} \cdot \mathbf{r}_{S \mathrm{i}}} \rho_{S \mathrm{i} \sigma},
$
where $N_c$ is the number of unit cells in the system and $\mathbf{r}_{S \mathrm{i}}$
is the position of the lattice site $S \in \mathrm{i}$, and performing a basis 
transformation from the sublattice to the band basis through
$
c_{S \mathbf{k} \sigma}^\dagger = \sum_{n} 
n_{S \mathbf{k} \sigma}^* c_{n \mathbf{k} \sigma}^\dagger,
$
we obtain
\begin{align}
\rho_{S \mathbf{q} \sigma} = \frac{1}{\sqrt{N_c}}\sum_{n m \mathbf{k}}
n^*_{S,\mathbf{k-q},\sigma} m_{S \mathbf{k} \sigma}
c_{n, \mathbf{k-q}, \sigma}^\dagger c_{m \mathbf{k} \sigma}.
\end{align}
Here, we also used 
$
\sum_{\mathrm{i}} e^{i(\mathbf{q-q'})\cdot \mathbf{r}_{S \mathrm{i}}} 
= N_c \delta_{\mathbf{q}\mathbf{q'}}.
$
Note that 
$
\rho_{S \mathbf{q} \sigma}^\dagger = \rho_{S, \mathbf{-q}, \sigma},
$
and
$
\rho_\mathbf{q} = \sum_{S \sigma} \rho_{S \mathbf{q} \sigma}
$
for the density operator and
$
\rho_{s\mathbf{q}} = \sum_{S} (\rho_{S \mathbf{q} \uparrow} 
- \rho_{S \mathbf{q} \downarrow})
$
for the spin-density operator. As a result, the multi-orbital Hubbard model 
$
\mathcal{H} = \mathcal{H}_0 + \mathcal{H}_I
$
with onsite attractive interactions $U \ge 0$, i.e.,
$
\mathcal{H}_0 = -\sum_{S \mathrm{i}; S' \mathrm{i}'; \sigma} 
t_{S \mathrm{i}; S'\mathrm{i}'}^\sigma 
c_{S \mathrm{i} \sigma}^\dagger c_{S' \mathrm{i}' \sigma}
$
and
$
\mathcal{H}_I = -U \sum_{S \mathrm{i}} \rho_{S \mathrm{i} \uparrow} 
\rho_{S \mathrm{i} \downarrow},
$
can be equivalently written as
\begin{align}
\label{eqn:ham0}
\mathcal{H}_0 &= \sum_{S S' \mathbf{k} \sigma} c_{S \mathbf{k} \sigma}^\dagger 
(h_{\mathbf{k} \sigma}^{SS'} - \mu \delta_{SS'}) c_{S' \mathbf{k} \sigma},
\\
\mathcal{H}_I &= - U\sum_{S \mathbf{q}} 
\rho_{S \mathbf{q} \uparrow} \rho_{S, \mathbf{-q}, \downarrow},
\label{eqn:hamI}
\end{align}
in momentum space. Here, the hopping parameters 
$
t_{S \mathrm{i}; S' \mathrm{i}'}^\sigma
$ 
describe tunneling of a spin-$\sigma$ particle from the sublattice site 
$S' \in \mathrm{i}'$ to the sublattice site $S \in \mathrm{i}$,
$
h_{\mathbf{k} \sigma}^{SS'} = -\frac{1}{N_c}\sum_{\mathrm{i} \mathrm{i}'} 
t_{S\mathrm{i};S'\mathrm{i}'}^\sigma
e^{\mathrm{i} \mathbf{k} \cdot \mathbf{r}_{S\mathrm{i};S'\mathrm{i}'}}
$
with 
$
\mathbf{r}_{S\mathrm{i};S'\mathrm{i}'} = \mathbf{r}_{S'\mathrm{i}'} 
- \mathbf{r}_{S\mathrm{i}}
$
denoting the relative position, and $\mu$ is the chemical potential.

In order to make analytical progress with the f-sum rule, 
we first note that 
$
[\rho_{S \mathbf{q} \sigma}, \rho_{S' \mathbf{q'} \sigma'}] = 0
$
leading to
$
[\rho_{S \mathbf{q} \sigma}, \mathcal{H}_I] = 0.
$
Then, after some simple algebra~\footnote{
For instance, by making use of the commutator relation
$
[\rho_{S \mathbf{q} \sigma}, \mathcal{H}_0] = \frac{1}{\sqrt{N_c}} \sum_{n m \mathbf{k}}
(\varepsilon_{m \mathbf{k} \sigma} - \varepsilon_{n, \mathbf{k-q}, \sigma})
n_{S, \mathbf{k-q}, \sigma}^* m_{S \mathbf{k} \sigma}
c_{n, \mathbf{k-q},\sigma}^\dagger c_{m \mathbf{k} \sigma},
$
where the non-interacting Hamiltonian $\mathcal{H}_0$ is given in 
Eq.~(\ref{eqn:ham0}), one can show that 
$
[\rho_{S' \mathbf{q'} \sigma'}, [\rho_{S \mathbf{q} \sigma}, \mathcal{H}_0] 
= \frac{\delta_{\sigma \sigma'}}{N_c} \sum_{nml \mathbf{k}} 
(\varepsilon_{m \mathbf{k} \sigma} - \varepsilon_{n, \mathbf{k-q}, \sigma})
n_{S, \mathbf{k-q}, \sigma}^* m_{S \mathbf{k} \sigma}
(
l_{S',\mathbf{k-q-q'},\sigma}^* n_{S',\mathbf{k-q},\sigma} 
c_{l,\mathbf{k-q-q'},\sigma}^\dagger c_{m \mathbf{k} \sigma} 
-
m_{S' \mathbf{k} \sigma}^* l_{S',\mathbf{k+q'},\sigma} 
c_{n,\mathbf{k-q},\sigma}^\dagger c_{l, \mathbf{k+q'}, \sigma}
)
$
},
we find
\begin{align}
&[\rho_\mathbf{q}^\dagger, [\rho_\mathbf{q}, \mathcal{H}]] = 
\frac{1}{N_c} \sum_{nml\mathbf{k}\sigma}
(\varepsilon_{m \mathbf{k} \sigma} - \varepsilon_{n, \mathbf{k-q}, \sigma})
\langle n_{\mathbf{k-q},\sigma} | m_{\mathbf{k} \sigma} \rangle
\nonumber \\
&\times \big(
\langle l_{\mathbf{k} \sigma} | n_{\mathbf{k-q}, \sigma} \rangle
c_{l \mathbf{k} \sigma}^\dagger c_{m \mathbf{k} \sigma}
- \langle m_{\mathbf{k} \sigma} | l_{\mathbf{k-q}, \sigma} \rangle
c_{n, \mathbf{k-q}, \sigma}^\dagger c_{l, \mathbf{k-q}, \sigma}
\big),
\end{align}
where 
$
\langle n_{\mathbf{k}\sigma} | m_{\mathbf{k'} \sigma} \rangle
= \sum_S n_{S \mathbf{k} \sigma}^* m_{S \mathbf{k'} \sigma}
$
and
$
\langle n_{\mathbf{k}\sigma} | m_{\mathbf{k} \sigma} \rangle 
= \delta_{nm}.
$
If only intraband averages are nonzero with respect to the ground state, 
i.e., 
$
\langle c_{n \mathbf{k} \sigma}^\dagger c_{m \mathbf{k} \sigma} \rangle = 
\langle c_{n \mathbf{k} \sigma}^\dagger c_{n \mathbf{k} \sigma} \rangle \delta_{nm},
$
then we find
\begin{align}
\label{eqn:fscbcs}
&\langle [\rho_\mathbf{q}^\dagger, [\rho_\mathbf{q}, \mathcal{H}]] \rangle
= - \frac{1}{N_c} \sum_{ij n \mathbf{k} \sigma} 
(M_{n\mathbf{k}\sigma}^{-1})_{ij} q_i q_j
\langle c_{n \mathbf{k} \sigma}^\dagger c_{n \mathbf{k} \sigma} \rangle,
\\
&(M_{n\mathbf{k}\sigma}^{-1})_{ij} = 
\ddot{\varepsilon}_{n\mathbf{k}\sigma}^{ij}
- \sum_{m \ne n} (\varepsilon_{n \mathbf{k} \sigma} - \varepsilon_{m \mathbf{k} \sigma})
g_{ij}^{n m\mathbf{k} \sigma},
\label{eqn:emt}
\end{align}
in the low-$\mathbf{q}$ limit up to second-order in $\mathbf{q}$, 
where 
$
\ddot{\varepsilon}_{n\mathbf{k}\sigma}^{ij} 
= \partial^2 \varepsilon_{n\mathbf{k}\sigma} / (\partial k_i \partial k_j)
$
~\footnote{
In general, the low-$\mathbf{q}$ limit can be written as
$
[\rho_\mathbf{q}^\dagger, [\rho_\mathbf{q}, \mathcal{H}]] 
= - \frac{1}{N_c} \sum_{ij nm \mathbf{k} \sigma} 
(M_{nm\mathbf{k}\sigma}^{-1})_{ij} q_i q_j
c_{n \mathbf{k} \sigma}^\dagger c_{m \mathbf{k} \sigma},
$
up to second-order in $\mathbf{q}$, where the coefficients for the 
inverse effective-mass tensor are given by
$
(M_{nm\mathbf{k}\sigma}^{-1})_{ij} = 
\ddot{\varepsilon}_{n\mathbf{k}\sigma}^{ij} \delta_{nm}
+2 (\dot{\varepsilon}_{n \mathbf{k} \sigma}^i - \dot{\varepsilon}_{m \mathbf{k} \sigma}^i)
\langle \dot{n}_{\mathbf{k} \sigma}^{j} | m_{\mathbf{k} \sigma} \rangle
+(\varepsilon_{n \mathbf{k} \sigma} - \varepsilon_{m \mathbf{k} \sigma})
(\langle \ddot{n}_{\mathbf{k} \sigma}^{ij} | m_{\mathbf{k} \sigma} \rangle
+ \langle \dot{n}_{\mathbf{k} \sigma}^i | \dot{m}_{\mathbf{k} \sigma}^j \rangle)
+ \sum_l (2\varepsilon_{l \mathbf{k} \sigma} - \varepsilon_{n \mathbf{k} \sigma} 
- \varepsilon_{m \mathbf{k} \sigma})
\langle \dot{l}_{\mathbf{k} \sigma}^i | m_{\mathbf{k} \sigma} \rangle
\langle n_{\mathbf{k} \sigma} | \dot{l}_{\mathbf{k} \sigma}^j \rangle
$
}.
Equation~(\ref{eqn:emt}) is the so-called effective-mass theorem
for the $n$th Bloch band in a multiband setting~\cite{iskin19b}.
The matrix elements $(M_{n\mathbf{k}\sigma}^{-1})_{ij}$ of the 
inverse effective-mass tensor are given by 
$
\langle n_{\mathbf{k}\sigma} | \ddot{\mathbf{h}}_{\mathbf{k} \sigma}^{ij} 
| n_{\mathbf{k} \sigma} \rangle
$
~\cite{onishi24a, verma25, yu24}. Here,
\begin{align}
g_{ij}^{n m\mathbf{k}\sigma} = 2\mathrm{Re} \left[ 
\langle \dot{n}_{\mathbf{k} \sigma}^i | m_{\mathbf{k}\sigma} \rangle
\langle m_{\mathbf{k} \sigma} | \dot{n}_{\mathbf{k} \sigma}^j \rangle
\right]
\end{align}
is the band-resolved quantum-metric tensor, where $\mathrm{Re}$ 
denotes the real part of the expression and 
$
|\dot{n}_{\mathbf{k} \sigma}^i \rangle = 
\partial |n_{\mathbf{k} \sigma} \rangle / \partial k_i.
$
The quantum-metric tensor of the $n$th Bloch band can be written as
$
g_{ij}^{n\mathbf{k} \sigma}  = \sum_{m\ne n} g_{ij}^{n m\mathbf{k} \sigma}
$
~\cite{Provost80, berry84, resta11}. 
In Sec.~\ref{sec:gc}, where we analyze multiband BCS superconductors, 
we reveal that the quantum-geometric origin of the superfluid-weight 
tensor, along with related observables such as the effective-mass tensor 
of Cooper pairs, coherence length and low-energy structure factors, 
can be traced back to Eq.~(\ref{eqn:emt}) in the strong-coupling limit.

Whenever a physical quantity is related to the quantum metric, it is
important to carefully consider how the metric depends on the positions
of the orbitals. In particular, Eq.(\ref{eqn:emt}) explicitly reflects 
this dependence through the term $g_{ij}^{n m\mathbf{k} \sigma}$. 
Since this equation appears in other physical contexts as 
well~\cite{onishi24a, verma25, yu24}, further investigation is required 
to fully understand the implications of this dependence. That said, in 
the rest of this paper, we restrict our analysis to models with 
time-reversal symmetry and uniform pairing. In these cases, the orbital 
positions are fixed by symmetry at high-symmetry points, simplifying 
the situation.

\section{Linear-response theory for Multiband BCS superconductors}
\label{sec:lrt}

To make further analytical progress, we consider a generic 
multi-ortbital Hubbard model that exhibits time-reversal symmetry and 
uniform BCS pairing across the underlying sublattices within a unit cell.
The former condition requires
$
\mathbf{h}_{-\mathbf{k}, \downarrow}^* = \mathbf{h}_{\mathbf{k} \uparrow}
= \mathbf{h}_{\mathbf{k}},
$
which leads to
$
n_{S, -\mathbf{k}, \downarrow}^* = n_{S \mathbf{k} \uparrow} 
= n_{S \mathbf{k}} 
$
and 
$
\varepsilon_{n, -\mathbf{k}, \downarrow} = \varepsilon_{n \mathbf{k} \uparrow} 
= \varepsilon_{n \mathbf{k}}.
$
Time-reversal symmetry, along with inversion symmetry, guarantees that 
for every spin-$\downarrow$ electron in the state $(n, \mathbf{-k})$,
there exists a corresponding spin-$\uparrow$ electron in the state 
$(n, \mathbf{k})$. Consequently, the condensation of spin-singlet Cooper 
pairs with zero center-of-mass momentum forms the ground state, as in 
the usual BCS theory. Under these assumptions, there is no interband 
pairing and only intraband averages remain nonzero with respect to the 
BCS ground state~\cite{liang17, iskin24a}. For instance,
$
\langle c_{n \mathbf{k} \sigma}^\dagger c_{m \mathbf{k} \sigma} \rangle
= v_{n \mathbf{k}}^2 \delta_{nm},
$
where the coherence factor is given by
$
v_{n \mathbf{k}}^2 = 1/2 - \xi_{n \mathbf{k}}/(2E_{n \mathbf{k}}),
$
with
$
\xi_{n \mathbf{k}} = \varepsilon_{n \mathbf{k}} - \mu
$
and
$
E_{n \mathbf{k}} = \sqrt{\xi_{n \mathbf{k}}^2 + \Delta_0^2}
$
denoting the dispersion of quasiparticles. 
Here, $\Delta_0$ is the superconducting order parameter for pairing.
Thus, by combining Eqs.~(\ref{eqn:fsc}) and~(\ref{eqn:fscbcs}), we obtain the 
f-sum rule for the multiband BCS superconductors,
\begin{align}
2\int_0^{\infty} d\omega \omega S(\mathbf{q}, \omega) = 
\frac{1}{N_c} \sum_{ij n\mathbf{k}} (M_{n\mathbf{k}}^{-1})_{ij} q_i q_j 
\bigg(1 - \frac{\xi_{n\mathbf{k}}}{E_{n\mathbf{k}}}\bigg),
\label{eqn:fsrbcs}
\end{align}
which is valid in the low-$\mathbf{q}$ limit up to second-order in $\mathbf{q}$.
Here, the effective-mass tensor of Bloch bands satisfies
$
\mathbf{M}_{n,\mathbf{-k},\downarrow} = \mathbf{M}_{n\mathbf{k}\uparrow} 
= \mathbf{M}_{n\mathbf{k}}.
$
It is pleasing to see that Eq.~(\ref{eqn:fsrbcs}) recovers the well-known 
textbook result for the continuum model~\cite{nozieres1999theory, 
combescot06, guo13, he16}.

\subsection{BCS-BEC crossover at $T = 0$}
\label{sec:bbc}

Thanks to time-reversal symmetry and the uniform pairing condition, 
we define the BCS expectation value
$
\Delta_{S \mathrm{i}} = U \langle 
c_{S \mathrm{i} \uparrow} c_{S \mathrm{i} \downarrow}
\rangle
$
as $\Delta_0$ for every sublattice site $S$ within any unit cell $\mathrm{i}$. 
Then, within the zero-temperature BCS mean-field theory~\cite{nozieres1985bose}, 
the parameters $\Delta_0$ and $\mu$ are determined by the self-consistent 
solutions of~\cite{iskin18c, iskin24a}
\begin{align}
\label{eqn:op}
1 &= \frac{U}{N} \sum_{n\mathbf{k}} \frac{1}{2E_{n\mathbf{k}}}, \\
F &= 1 - \frac{1}{N} \sum_{n\mathbf{k}} 
\frac{\xi_{n\mathbf{k}}}{E_{n\mathbf{k}}},
\label{eqn:filling}
\end{align}
where $N = N_b N_c$ is the total number of lattice sites, with $N_b$ being 
the number of sublattices in a unit cell. Here, the filling 
$0 \le 
F = \frac{1}{N}\sum_{n \mathbf{k} \sigma} 
\langle c_{n \mathbf{k} \sigma}^\dagger c_{n \mathbf{k} \sigma} \rangle
\le 2$
corresponds to the total number of particles per site. Furthermore, the 
mean-field expression for the filling of condensed particles is 
given by~\cite{leggett, iskin18c, iskin24a, iskin25a} 
\begin{align}
\label{eqn:Fc}
F_c = \frac{1}{N} \sum_{n\mathbf{k}} \frac{\Delta_0^2}
{2E_{n\mathbf{k}}^2}.
\end{align}
It is known that Eqs.~(\ref{eqn:op}),~(\ref{eqn:filling}) and~(\ref{eqn:Fc}) 
are sufficient at $T = 0$ to provide a qualitative description of the 
BCS-BEC crossover for all values of $U$.

\subsection{Goldstone modes}
\label{sec:gm}

The dispersions $\omega_\mathbf{q}$ of the collective modes are 
determined by the poles of the fluctuation propagator, as discussed 
in Appendix~\ref{sec:gf}~\cite{engelbrecht97, he16}.
In the particular case of multiband BCS superconductors with uniform 
pairing fluctuations~\cite{iskin24a}, and assuming they are below the 
quasiparticle continuum, the low-energy collective modes are determined by
$
I^{11}_{\mathbf{q} \omega_\mathbf{q}} I^{22}_{\mathbf{q} \omega_\mathbf{q}}
-\omega_\mathbf{q}^2 (I^{12}_{\mathbf{q} \omega_\mathbf{q}})^2 = 0,
$
where
\begin{align}
I^{11}_{\mathbf{q} \omega} &= \frac{2N_b}{U} + \frac{1}{N_c} \sum_{nm\mathbf{k}} 
\frac{E + E'}{EE'} \frac{EE' + \xi \xi' - \Delta_0^2}{\omega^2 - (E+E')^2} 
Z, \\
I^{22}_{\mathbf{q} \omega} &= \frac{2N_b}{U} + \frac{1}{N_c} \sum_{nm\mathbf{k}} 
\frac{E + E'}{EE'} \frac{EE' + \xi \xi' + \Delta_0^2}{\omega^2 - (E+E')^2} 
Z, \\
I^{12}_{\mathbf{q} \omega} &= \frac{1}{N_c} \sum_{nm\mathbf{k}} 
\bigg(\frac{\xi}{E} + \frac{\xi'}{E'}\bigg) \frac{Z}{\omega^2 - (E+E')^2}. 
\end{align}
Here, we use the shorthand notations $\xi = \xi_{n \mathbf{k}}$, 
$\xi' = \xi_{m, \mathbf{k-q}}$, $E = E_{n \mathbf{k}}$, 
$E' = E_{m, \mathbf{k-q}}$, and
$
Z = |\langle n_\mathbf{k} | m_\mathbf{k-q} \rangle|^2
$
for convenience. 
In the low-$\mathbf{q}$ limit, $I^{11}_{\mathbf{q} \omega}$ describes the 
amplitude-amplitude degrees of freedom of the fluctuation propagator, 
$I^{22}_{\mathbf{q} \omega}$ describes the phase-phase degrees of freedom, 
and $I^{12}_{\mathbf{q} \omega}$ describes the amplitude-phase coupling. 
Using Eq.~(\ref{eqn:op}), we observe that
$
I^{22}_{\mathbf{0} \omega} = \frac{1}{N_c} \sum_{n \mathbf{k}}
[4E_{n \mathbf{k}}/(\omega^2 - 4E_{n \mathbf{k}}^2) + 1/E_{n\mathbf{k}}]
$
vanishes as $\omega \to 0$, indicating that the low-frequency phase mode is 
always gapless, which allows us to identify it as the in-phase Goldstone mode. 
Similarly, $I^{11}_{\mathbf{0} \omega}$ vanishes as $\omega \to 2\Delta_0$, 
again due to Eq.~(\ref{eqn:op}), suggesting that the amplitude mode is 
gapped with energy $2\Delta_0$. This mode is identified as the Higgs mode. 
While the out-of-phase Leggett modes are not included in this calculation 
or in the rest of this paper for the sake of simplicity (i.e., they are 
not analytically tractable in general), this omission is justified in 
Sec.~\ref{sec:gc}, where it is shown that the Goldstone modes alone 
exhaust the f-sum rule in the strong-coupling limit.

To determine the collective phonon modes, we expand
$
I^{11}_{\mathbf{q} \omega} = 2\bar{A} + 2\sum_{ij} \bar{C}_{ij} q_i q_j 
- 2\bar{D}\omega^2,
$
$
I^{22}_{\mathbf{q} \omega} = 2\sum_{ij} \bar{Q}_{ij} q_i q_j 
- 2\bar{R}\omega^2,
$
and
$
I^{12}_{\mathbf{q} \omega}= -2\bar{B}
$
up to second order in $\mathbf{q}$ and $\omega$~\cite{engelbrecht97, iskin24a}. 
The coefficients 
$
\bar{A} = \frac{1}{N_c} \sum_{n \mathbf{k}} \Delta_0^2/(2E_{n \mathbf{k}}^3)
= 4\Delta_0^2 \bar{R},
$
$
\bar{B} = \frac{1}{N_c} \sum_{n \mathbf{k}} \xi_{n \mathbf{k}}/(4E_{n \mathbf{k}}^3)
$
and
$
\bar{D} = \frac{1}{N_c} \sum_{n \mathbf{k}} \xi_{n \mathbf{k}}^2/(8E_{n \mathbf{k}}^5)
$
are simply sums of their conventional single-band counterparts.
However, the kinetic expansion coefficients
$
\bar{Q}_{ij} = \bar{Q}_{ij}^\mathrm{intra} + \bar{Q}_{ij}^\mathrm{inter}
$
and 
$
\bar{C}_{ij} = \bar{C}_{ij}^\mathrm{intra} + \bar{C}_{ij}^\mathrm{inter}
$
have both intraband and interband contributions, where
\begin{align}
\bar{Q}_{ij}^\mathrm{intra} &= \frac{1}{N_c} \sum_{n \mathbf{k}} 
\frac{\dot{\xi}_{n\mathbf{k}}^i \dot{\xi}_{n\mathbf{k}}^j}
{8 E_{n\mathbf{k}}^3}, 
\label{eqn:Qintra}
\\
\bar{Q}_{ij}^\mathrm{inter} &= \frac{1}{N_c} \sum_{n, m\ne n, \mathbf{k}} 
\frac{(\xi_{n\mathbf{k}}-\xi_{m\mathbf{k}})^2}
{8 E_{n\mathbf{k}} E_{m\mathbf{k}} (E_{n\mathbf{k}}+E_{m\mathbf{k}})}
g^{nm\mathbf{k}}_{ij}, 
\label{eqn:Qinter}
\\
\bar{C}_{ij}^\mathrm{intra} &= \frac{1}{N_c} \sum_{n \mathbf{k}} 
\left( 1 - \frac{5\Delta_0^2 \xi_{n\mathbf{k}}^2}{E_{n\mathbf{k}}^4} \right)
\frac{\dot{\xi}_{n\mathbf{k}}^i \dot{\xi}_{n\mathbf{k}}^j}{8 E_{n\mathbf{k}}^3}, 
\label{eqn:Cintra}
\\
\bar{C}_{ij}^\mathrm{inter} &= 
- \frac{1}{N_c} \sum_{n \mathbf{k}} 
\frac{\Delta_0^2} {4 E_{n\mathbf{k}}^3} g^{n\mathbf{k}}_{ij} \nonumber \\
&+ \frac{1}{N_c} \sum_{n, m\ne n, \mathbf{k}} 
\frac{(\xi_{n\mathbf{k}}-\xi_{m\mathbf{k}})^2 + 4\Delta_0^2}
{8 E_{n\mathbf{k}} E_{m\mathbf{k}} (E_{n\mathbf{k}}+E_{m\mathbf{k}})}
g^{nm\mathbf{k}}_{ij}.
\label{eqn:Cinter}
\end{align}
The geometric terms follow from the low-$\mathbf{q}$ expansion of the Bloch factor 
$
Z = |\langle n_\mathbf{k} | m_\mathbf{k-q} \rangle|^2,
$
which can be expressed as
\begin{align}
Z = \delta_{nm} - \frac{1}{2} \sum_{ij} [g_{ij}^{n\mathbf{k}} \delta_{nm} 
+ g_{ij}^{nm\mathbf{k}}(\delta_{nm} - 1)] q_i q_j,
\label{eqn:gnm}
\end{align}
up to second-order in $\mathbf{q}$. Here,
$
g_{ij}^{n\mathbf{k}}  = \sum_{m\ne n} g_{ij}^{n m\mathbf{k}}
$
and
$
g_{ij}^{n m\mathbf{k}} = 2\mathrm{Re} [ 
\langle \dot{n}_\mathbf{k}^i | m_\mathbf{k} \rangle
\langle m_\mathbf{k} | \dot{n}_\mathbf{k}^j \rangle].
$
The separation ensures that the coefficients $Q_{ij}^\mathrm{intra}$ 
and $C_{ij}^\mathrm{intra}$ remain simple sums of their conventional 
single-band counterparts. 
Note that $\bar{Q}_{ij}^\mathrm{inter}$ and $\bar{C}_{ij}^\mathrm{inter}$
do not have any contribution from the band touchings, 
e.g., the first term of Eq.~(\ref{eqn:Cinter}) cancel those 
touching contributions from the second term whenever 
$\xi_{n\mathbf{k}} = \xi_{m\mathbf{k}}$ for any $n \ne m$. 
As a result, the Goldstone modes in the low-$\mathbf{q}$ limit are 
given by~\cite{engelbrecht97, iskin24a}
\begin{align}
(\omega_\mathbf{q}^G)^2 = \sum_{ij} 
\frac{\bar{Q}_{ij}}{\bar{R}+ \bar{B}^2/\bar{A}} q_i q_j.
\end{align}
These collective modes are related to the superfluid-weight tensor 
$\mathcal{D}_{ij}$ through
$
\mathcal{D}_{ij} = \frac{8N_c \Delta_0^2}{V} \bar{Q}_{ij}
$
at $T = 0$~\cite{iskin24a}. On the other hand, $\bar{C}_{ij}$ is 
related to the zero-temperature coherence length $\xi_\mathrm{0}$ through
$
(\xi_\mathrm{0}^2)_{ij} = \bar{C}_{ij}/\bar{A}
$
~\cite{iskin24c}. This length scale can, in turn, be connected to the 
Ginzburg–Landau coherence length in the dilute limit~\cite{chen23, iskin23}.

\subsection{Response functions and f-sum rule}
\label{sec:rf}

In this section, following Ref.~\cite{he16}, we derive the density and 
spin response functions for multiband BCS superconductors using the 
imaginary-time functional path-integral formalism. 
This approach allows us to calculate the density-density and 
spin-spin correlation functions by introducing the corresponding source 
terms. Consequently, after a lengthy but straightforward calculation, 
the density and spin response functions can be expressed as
\begin{align}
\label{eqn:Xnn}
\mathcal{X}(\mathbf{q}, \omega) &= \mathcal{X}_s(\mathbf{q}, \omega) 
+ 2\Delta_0^2 B_{\mathbf{q} \omega} + \mathcal{X}_{c}(\mathbf{q}, \omega),
\\
\label{eqn:Xss}
\mathcal{X}_s(\mathbf{q}, \omega) &= \frac{1}{N_c} \sum_{nm\mathbf{k}} 
\frac{E + E'}{EE'} \frac{EE' - \xi \xi' - \Delta_0^2}{\omega^2 - (E+E')^2} 
Z,
\\
\label{eqn:Xcc}
\mathcal{X}_{c}(\mathbf{q}, \omega) &= \frac{\Delta_0^2 C_{\mathbf{q} \omega}}
{\omega^2 (I^{12}_{\mathbf{q} \omega})^2
- I^{11}_{\mathbf{q} \omega} I^{22}_{\mathbf{q} \omega}},
\end{align}
where
$
C_{\mathbf{q} \omega} = A^2_{\mathbf{q} \omega} I^{22}_{\mathbf{q} \omega}
+ \omega^2 B^2_{\mathbf{q} \omega} I^{11}_{\mathbf{q} \omega}
- 2\omega^2 A_{\mathbf{q} \omega} B_{\mathbf{q} \omega} I^{12}_{\mathbf{q} \omega},
$
and
$
A_{\mathbf{q} \omega} = \frac{1}{N_c} \sum_{nm\mathbf{k}} 
\frac{E + E'}{EE'} \frac{\xi + \xi'}{\omega^2 - (E+E')^2} 
Z
$
and
$
B_{\mathbf{q} \omega} = \frac{1}{N_c} \sum_{nm\mathbf{k}} 
\frac{E + E'}{EE'} \frac{Z}{\omega^2 - (E+E')^2}.
$
Similar to the well-known continuum case~\cite{combescot06, guo13, he16}, 
while the density response couples to the collective modes,
the spin response does not. In Sec.~\ref{sec:gm}, we show that the 
collective-mode contribution $\mathcal{X}_{c}(\mathbf{q}, \omega)$ 
plays a crucial role in the proper description of the BEC regime 
in the BCS-BEC crossover
~\footnote{
We also note that, if we ignore the collective-mode contribution 
$\mathcal{X}_{c}(\mathbf{q}, \omega)$ and set it to zero, the 
compressibility sum rule 
$
2 \int_0^{\infty} \frac{d\omega}{\omega} S(\mathbf{q}, \omega) 
= \frac{1}{N_c} \sum_{n \mathbf{k}} \Delta_0^2/E_{n \mathbf{k}}^3
= N_b \partial F/\partial \mu
$
is satisfied at the mean-field level, where $N_b F$ is the particle
filling per unit cell
}. 
Thus, the dynamic SSF and its f-sum rule are simply given by
\begin{align}
& S_s(\mathbf{q}, \omega) = \frac{1}{N_c} \sum_{nm\mathbf{k}} 
\frac{EE' - \xi \xi' - \Delta_0^2}{2EE'} Z \delta(\omega - E - E'),
\\ 
& \int_0^{\infty} d\omega \omega S_s(\mathbf{q}, \omega) 
= \frac{1}{N_c} \sum_{nm\mathbf{k}} \frac{E+E'}{2EE'}
(EE' - \xi \xi' - \Delta_0^2) Z.
\end{align}
Since the SSF is immune to the collective modes, these  
expressions are valid for any $U$ or $\mathbf{q}$ within the 
linear-response theory. For instance, it is reassuring to verify 
that the latter is in perfect agreement with 
Eqs.~(\ref{eqn:fsrs}) and~(\ref{eqn:fsrbcs}) in the low-$\mathbf{q}$ 
limit, i.e., after using integration by parts
$
\sum_{\mathbf{k}} \xi_{n \mathbf{k}} \ddot{\xi}_{n \mathbf{k}}^{ij}/E_{n \mathbf{k}}
= - \Delta_0^2 \sum_{\mathbf{k}} \dot{\xi}_{n\mathbf{k}}^i \dot{\xi}_{n\mathbf{k}}^j
/E_{n \mathbf{k}}^3
$
along with the symmetries
$
g_{ij}^{nm\mathbf{k}} = g_{ji}^{mn\mathbf{k}},
$
$
g_{ij}^{n\mathbf{k}} = g_{ji}^{n\mathbf{k}}
$
and
$
\sum_{n, m \ne n, \mathbf{k}}
(\varepsilon_{n\mathbf{k}} - \varepsilon_{m\mathbf{k}}) 
g^{nm\mathbf{k}}_{ij} = 0.
$
The static SSF $S_s(\mathbf{q})$ is determined by
$
\int_0^{\infty} d\omega S_s(\mathbf{q}, \omega) 
= \frac{1}{N_c} \sum_{nm\mathbf{k}} (EE' - \xi \xi' - \Delta_0^2) Z/(2EE')
$
~\footnote{
The low-$\mathbf{q}$ expansion of the static SSF $S_s(\mathbf{q})$ 
can be determined as
$
\int_0^\infty d\omega S_s(\mathbf{q}, \omega) = 
\sum_{ij} \frac{q_i q_j}{N_c} 
[ 
\Delta_0^2 \sum_{n \mathbf{k}} 
\dot{\xi}_{n\mathbf{k}}^i \dot{\xi}_{n\mathbf{k}}^j / (4E_{n \mathbf{k}}^4)
+ \sum_{n, m\ne n, \mathbf{k}} 
(E_{n \mathbf{k}} E_{m \mathbf{k}} - \xi_{n \mathbf{k}} \xi_{m \mathbf{k}} - \Delta_0^2)
g_{ij}^{nm \mathbf{k}}/(4E_{n \mathbf{k}} E_{m \mathbf{k}})
]
$
}.
Note also that 
$
4\mathcal{X}_{\uparrow\downarrow}(\mathbf{q}, \omega) 
= 2\Delta_0^2 B_{\mathbf{q} \omega} 
+ \mathcal{X}_{c}(\mathbf{q}, \omega)
$
corresponds to the spin-resolved response function
~\footnote{
Note that the $B_{\mathbf{q} \omega}$-term has the following contributions 
to the f-sum rule
$
-\frac{1}{\pi}
\int_0^\infty d\omega \omega \mathrm{Im} B_{\mathbf{q}, \omega+i0^+}
= 2N_b/U
$
and static DSF
$
-\frac{1}{\pi}
\int_0^\infty d\omega \mathrm{Im} B_{\mathbf{q}, \omega+i0^+}
= \frac{1}{2N_c} \sum_{n \mathbf{k}} 1/E_{n \mathbf{k}}^2
- \sum_{ij} \frac{q_i q_j}{4N_c} 
[ 
\sum_{n \mathbf{k}} 
\dot{\xi}_{n\mathbf{k}}^i \dot{\xi}_{n\mathbf{k}}^j \xi_{n\mathbf{k}}^2/ E_{n \mathbf{k}}^6
- \sum_{n, m\ne n, \mathbf{k}} 
g_{ij}^{nm \mathbf{k}}/(E_{n \mathbf{k}} E_{m \mathbf{k}})
]
$
up to second-order in $\mathbf{q}$
}.

\subsection{Goldstone contribution}
\label{sec:gc}

Due to its coupling to the collective phonon modes, the dynamic DSF is 
quite complicated for high frequencies. However, in the range of 
frequencies $0 \le \omega < \Theta$, where 
$
\Theta = \min(E_{n\mathbf{k}} + E_{m,\mathbf{k-q}})
$ 
is determined by the onset of the quasiparticle continuum,
the imaginary parts of 
$\mathcal{X}_s(\mathbf{q}, \omega)$, $I_{\mathbf{q} \omega}^{11}$, 
$I_{\mathbf{q} \omega}^{22}$, $I_{\mathbf{q} \omega}^{12}$,
$A_{\mathbf{q} \omega}$ and $B_{\mathbf{q} \omega}$ 
vanish when $\omega \to \omega+i0^+$. As a result, the density response
function 
$
\mathcal{X}(\mathbf{q}, \omega) \to \mathcal{X}_{c}(\mathbf{q}, \omega)
$
is determined solely by the collective phase-mode contribution in this 
frequency interval~\cite{buchler04, mitchison16}, where we reexpress it as
$
\mathcal{X}_{c}(\mathbf{q}, \omega) = 
\frac{\Delta_0^2 C_{\mathbf{q} \omega}}{(I_{\mathbf{q} \omega}^{12})^2} 
\frac{1}{\omega^2 - \Omega^2_{\mathbf{q} \omega}}.
$
Here,
$
\Omega_{\mathbf{q} \omega} = \sqrt{I_{\mathbf{q} \omega}^{11}
I_{\mathbf{q} \omega}^{22}}/I_{\mathbf{q} \omega}^{12},
$
and the collective modes are determined by the condition
$
\omega_\mathbf{q} = \Omega_{\mathbf{q} \omega_\mathbf{q}}.
$
Thus,
$
S(\mathbf{q}, \omega) \to S_{c}(\mathbf{q}, \omega)
$
in this frequency interval, where the pole contribution
\begin{align}
S_{c}(\mathbf{q}, \omega) = \frac{\Delta_0^2 C_{\mathbf{q} \omega_\mathbf{q}^G}}
{2 \omega_\mathbf{q}^G (I_{\mathbf{q} \omega_\mathbf{q}^G}^{12})^2} 
\delta(\omega-\omega_\mathbf{q}^G)
\end{align}
corresponds to the Goldstone modes in the low-$\mathbf{q}$ limit.
Using the small $\mathbf{q}$ and $\omega$ expansions that are given 
in Sec.~\ref{sec:gm}, along with 
$
A_{\mathbf{0} 0} = -4 \bar{B}
$
and
$
B_{\mathbf{0} 0} = -4 \bar{R},
$
we find
$
C_{\mathbf{q} \omega_\mathbf{q}^G} = 32(\bar{A} \bar{R} + \bar{B}^2) 
\sum_{ij} \bar{Q}_{ij} q_i q_j,
$
which leads to
$
S_{c}(\mathbf{q}, \omega) = \frac{V}
{2 \omega_\mathbf{q}^G N_c} ( 1 + \bar{A} \bar{R}/\bar{B}^2)
\sum_{ij} \mathcal{D}_{ij} q_i q_j
\delta(\omega-\omega_\mathbf{q}^G). 
$
In particular, in the strong-coupling limit, one can show that 
$
\bar{A} \bar{R} \ll \bar{B}^2
$ 
and the superfluid-weight tensor can be written as
$
\mathcal{D}_{ij} = 4 \rho_{Cp} (M_{Cp}^{-1})_{ij},
$
where $\rho_{Cp} = N F_{Cp}/V$ is the density and $(M_{Cp}^{-1})_{ij}$ 
is the inverse effective-mass tensor of condensed Cooper pairs
~\footnote{
Recall that the superfluid-weight tensor is
$
\mathcal{D}_{ij} = \frac{8N_c \Delta_0^2}{V} \bar{Q}_{ij},
$
where the expansion coefficient 
$
\bar{Q}_{ij} = \bar{Q}_{ij}^\mathrm{intra} + \bar{Q}_{ij}^\mathrm{inter}
$ 
can alternatively be expressed as
$
\bar{Q}_{ij}^\mathrm{intra} = 
\frac{1}{8\Delta_0^2 N_c}\sum_{n \mathbf{k}} 
(1 - \xi_{n \mathbf{k}}/E_{n \mathbf{k}})
\ddot{\xi}_{n \mathbf{k}}^{ij},
$
since
$
\sum_{\mathbf{k}} \ddot{\xi}_{n \mathbf{k}}^{ij} = 0,
$
and
$
\bar{Q}_{ij}^\mathrm{inter} = 
\frac{1}{4N_c}\sum_{n \mathbf{k}} g_{ij}^{n \mathbf{k}}/E_{n \mathbf{k}}
-\frac{1}{2N_c}\sum_{n, m \ne n, \mathbf{k}} \xi_{m\mathbf{k}} 
g_{ij}^{nm \mathbf{k}}/[E_{m \mathbf{k}}(\xi_{n\mathbf{k}} + \xi_{m\mathbf{k}})].
$
In the strong-coupling limit, they can be approximated as
$
\bar{Q}_{ij}^\mathrm{intra} = 
\frac{1}{16N_c}\sum_{n \mathbf{k}} 
\ddot{\xi}_{n \mathbf{k}}^{ij}/\xi_{n\mathbf{k}}^2
$
and
$
\bar{Q}_{ij}^\mathrm{inter} = 
\frac{1}{4N_c}\sum_{n \mathbf{k}} g_{ij}^{n \mathbf{k}}/|\xi_{n \mathbf{k}}|
-\frac{1}{2N_c}\sum_{n, m \ne n, \mathbf{k}} g_{ij}^{nm \mathbf{k}}
/(|\xi_{n\mathbf{k}}| + |\xi_{m\mathbf{k}}|),
$
showing direct relation to the effective-mass tensor of 
Cooper pairs~\cite{iskin24a}
}.
Here, $F_{Cp} = F_c/2$ is the filling of condensed Cooper pairs per lattice site.
We have recently shown that $\mathbf{M}_{Cp}$ depends weakly on $F$, 
and it is essentially given by the effective-mass tensor 
$
(M_{2b}^{-1})_{ij} = \frac{2}{U N} \sum_\mathbf{k} \mathrm{Tr}
(\dot{\mathbf{h}}_\mathbf{k}^i \dot{\mathbf{h}}_\mathbf{k}^j)
$ 
of lowest-bound two-body bound states~\cite{iskin24c}.
Thus, in the low-$\mathbf{q}$ limit, the dynamic DSF is dominated by
the collective phonon excitations 
\begin{align}
S_{c}(\mathbf{q}, \omega) = 4 \frac{N_b F_{Cp}} 
{2\omega_\mathbf{q}^G} \sum_{ij} (M_{Cp}^{-1})_{ij} q_i q_j
\delta(\omega-\omega_\mathbf{q}^G),
\label{eqn:sfgm}
\end{align}
up to second-order in $\mathbf{q}$, representing the Goldstone mode of the
broken symmetry. Here, $N_b F_{Cp}$ is the filling of condensed Cooper 
pairs per unit cell. 

It is pleasing to see that Eq.~(\ref{eqn:sfgm}) recovers the well-known 
textbook result for the dynamic DSF of a weakly-interacting Bose gas 
in a continuum model, i.e., it coincides with the phonon 
contribution within the Bogoliubov approximation~\cite{nozieres1999theory}. 
However, note that the resultant density-response function and structure 
factors are equivalent to four times that of the Cooper pairs, e.g.,
$
\mathcal{X}(\mathbf{q}, \omega) \to 4 \mathcal{X}_{Cp}(\mathbf{q}, \omega),
$
since the density of Cooper pairs is 
$
\rho_{Cp}(\mathbf{q}, t) = \frac{1}{2}\rho(\mathbf{q}, t)
$
in the strong-coupling limit~\cite{buchler04}.
Thus, since $\Theta  \to U \gg t$ in this limit, it is important to 
highlight that the Goldstone contribution alone exhausts the f-sum 
rule of the Cooper pairs~\cite{buchler04, mitchison16}, i.e., 
\begin{align}
2\int_0^{\Theta} d\omega \omega^\ell S_{c}(\mathbf{q}, \omega) = 
4 \frac{N_b F_{Cp}}{(\omega_\mathbf{q}^G)^{1-\ell}} \sum_{ij} (M_{Cp}^{-1})_{ij} q_i q_j
\label{eqn:sfbec}
\end{align}
when $\ell = 1$. Here, $\ell = -1$ corresponds to the compressibility 
sum rule, which is also exhausted, while $\ell = 0$ gives the static DSF,
which increases linearly with $|\mathbf{q}|$~\cite{nozieres1999theory}. 
Note that the f-sum rule of Eq.~(\ref{eqn:sfbec}) can also be written as 
$
\frac{V}{N_c} \sum_{ij} \mathcal{D}_{ij} q_i q_j,
$
which is consistent with the single-band result~\cite{monien92, denteneer93}.
These results indicate that the Goldstone modes are the only relevant 
degrees of freedom in the low-$\mathbf{q}$ limit at low frequencies. 
It is also reassuring to verify that Eq.~(\ref{eqn:sfbec}) is in perfect 
agreement with Eq.~(\ref{eqn:fsrbcs}) in the strong-coupling limit.
This is because, using
$
\Delta_0 = \frac{U}{2}\sqrt{F(2-F)}
$
and
$
\mu = -\frac{U}{2}(1-F)
$
in this limit, one can reduce
$
\mathcal{D}_{ij} = \frac{2F(2-F)}{UV} \sum_\mathbf{k} \mathrm{Tr}
(\dot{\mathbf{h}}_\mathbf{k}^i \dot{\mathbf{h}}_\mathbf{k}^j)
$
~\cite{iskin18c, iskin24a}, where $\mathrm{Tr}$ is the trace, and
$
\sum_{n\mathbf{k}} (M_{n\mathbf{k}}^{-1})_{ij} 
\big(1 - \xi_{n\mathbf{k}}/E_{n\mathbf{k}}\big)
\to -\frac{2F(2-F)}{U}
\sum_{n\mathbf{k}} (M_{n\mathbf{k}}^{-1})_{ij} \varepsilon_{n \mathbf{k}}
$
since
$
\sum_{n\mathbf{k}} (M_{n\mathbf{k}}^{-1})_{ij} = 0
$
in general and
$
\Delta_0^2/\sqrt{\mu^2+\Delta_0^2}^3 \to 2F(2-F/U.
$
Furthermore, we can use integration by parts to show that
$
\sum_{n\mathbf{k}} (M_{n\mathbf{k}}^{-1})_{ij} \varepsilon_{n \mathbf{k}}
= \sum_\mathbf{k} \mathrm{Tr}
(\ddot{\mathbf{h}}_\mathbf{k}^{ij} \mathbf{h}_\mathbf{k})
= - \sum_\mathbf{k} \mathrm{Tr}
(\dot{\mathbf{h}}_\mathbf{k}^i \dot{\mathbf{h}}_\mathbf{k}^j).
$
In other words, in the strong-coupling limit of a multiband BCS 
superconductor, the inverse effective-mass tensor of Cooper pairs 
can be written as
\begin{align}
(M_{Cp}^{-1})_{ij} &= -\frac{2}{U N} 
\sum_{n\mathbf{k}} (M_{n\mathbf{k}}^{-1})_{ij} \varepsilon_{n\mathbf{k}},
\label{eqn:Mcp}
\end{align}
for which $F_{Cp} = F(2-F)/4$~\cite{iskin24a}. Thus, the quantum-geometric 
origin of the pair mass can be traced all the way back to the effective-mass 
theorem for Bloch bands, as shown in Eq.~(\ref{eqn:emt})
~\footnote{
By reexpressing the superfluid-weight tensor as 
$
\mathcal{D}_{ij} = -\frac{2F(2-F)}{UV} 
\sum_{n\mathbf{k}} (M_{n\mathbf{k}}^{-1})_{ij} \varepsilon_{n\mathbf{k}},
$
we conclude that its quantum-geometric origin is also directly linked 
to Eq.~(\ref{eqn:emt}), i.e., to the effective-mass theorem for 
Bloch bands~\cite{iskin19b}, in the strong-coupling limit
}. See also Appendix~\ref{sec:emt}.
Lastly, the exhaustion of the f-sum rule by Eq.~(\ref{eqn:sfbec}) 
suggests that the remaining zeroth-order and second-order contributions 
to $\mathcal{X}(\mathbf{q}, \omega)$ from the high-frequency interval 
$\Theta \le \omega < \infty$ must necessarily vanish in this limit. 
See Ref.~\cite{guo13, he16} for an alternative interpretation 
in a continuum model~\footnote{
In Ref.~\cite{guo13}, we observed a crucial flaw in the argument used in 
the proof of the f-sum rule. Specifically, in Eq.~(53), they erroneously 
claim that the integral vanishes because the integrand is odd in $\omega$. 
However, by similar reasoning, one could argue that Eq.~(49) is even in 
$\omega$, which would imply that the integral of $\mathrm{Im}(\omega K^{00})$ 
should also vanish entirely. We thank C.-C. Chien for their email 
correspondence on this particular issue
}.

\section{Numerical Illustration}
\label{sec:ni}

To illustrate some of our main results, we choose a three-dimensional 
pyrochlore lattice with a four-point basis~\cite{wakefiel23, huang24},
described by
$
h^{SS}_{\mathbf{k} \sigma} = 0,
$
$
h^{AB}_{\mathbf{k} \sigma} = -2\bar{t} \cos\big(\frac{k_y+k_z}{4}a\big),
$
$
h^{AC}_{\mathbf{k} \sigma} = -2\bar{t} \cos\big(\frac{k_x+k_z}{4}a\big),
$
$
h^{AD}_{\mathbf{k} \sigma} = -2\bar{t} \cos\big(\frac{k_x+k_y}{4}a\big),
$
$
h^{BC}_{\mathbf{k} \sigma} = -2\bar{t} \cos\big(\frac{k_x-k_y}{4}a\big),
$
$
h^{BD}_{\mathbf{k} \sigma} = -2\bar{t} \cos\big(\frac{k_x-k_z}{4}a\big)
$
and
$
h^{CD}_{\mathbf{k} \sigma} = -2\bar{t} \cos\big(\frac{k_y-k_z}{4}a\big).
$
Here, $\bar{t}$ is the tight-binding hopping parameter between the 
nearest-neighbor sites, and $a$ is the side-length of the conventional 
simple-cubic cell~\cite{iskin24a}. The crystal structure forms a 
face-centered-cubic Bravais lattice, resulting in a truncated-octahedron-shaped 
Brillouin zone (BZ) with a side-length $\sqrt{2}\pi/a$, and the associated 
reciprocal space is such that
$
\sum_{\mathbf{k} \in \mathrm{BZ}} 1 = N_c.
$
Consequently, the Bloch spectrum consists of two dispersive bands
$
\varepsilon_{1\mathbf{k}\sigma} = -2\bar{t}(1 + \sqrt{1 + \alpha_\mathbf{k}})
$
and
$
\varepsilon_{2\mathbf{k}\sigma} = -2\bar{t}(1 - \sqrt{1 + \alpha_\mathbf{k}}),
$
where
$
\alpha_\mathbf{k} = \cos(k_x a/2) \cos(k_y a/2) + 
\cos(k_y a/2) \cos(k_z a/2) + \cos(k_x a/2) \cos(k_z a/2),
$
and two degenerate flat bands
$
\varepsilon_{3\mathbf{k}\sigma} = \varepsilon_{4\mathbf{k}\sigma} = 2\bar{t}.
$
Note that the flat bands touch $\varepsilon_{2\mathbf{k}\sigma}$ at $\mathbf{k} = 0$.  
In this paper, we set $\bar{t} \to -t$ and choose $t > 0$ as the unit of energy,  
so that the flat bands appear at the bottom of the Bloch spectrum.  
It turns out that the pyrochlore lattice is one of the simplest three-dimensional  
toy models that exhibit a flat band while preserving time-reversal symmetry  
and uniform pairing. For this reason, in our recent studies, we have also used 
this model to illustrate the quantum-geometric effects on the superfluid weight, 
collective modes, coherence length, pair size, and pair mass throughout 
the BCS-BEC crossover~\cite{iskin24a, iskin24c, iskin25a}. While those studies 
also focus on the pyrochlore lattice, they address distinct topics from the 
present work with only minor overlap.  

\begin{figure} [htb]
\includegraphics[width = 0.99\linewidth]{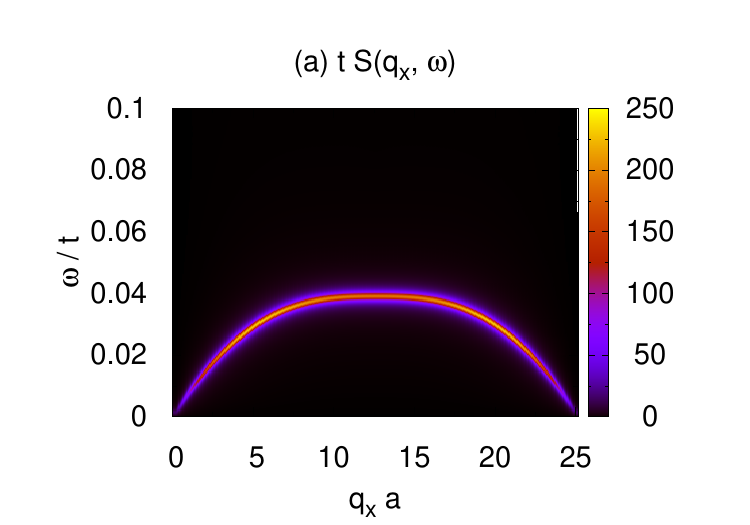}
\includegraphics[width = 0.99\linewidth]{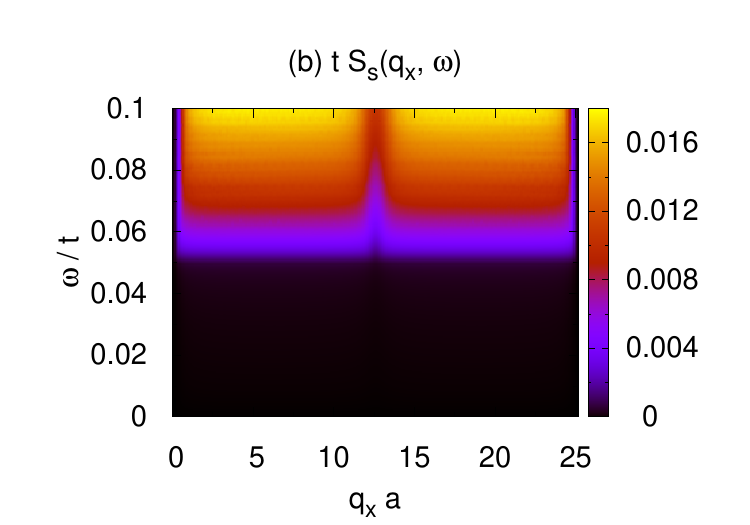}
\caption{\label{fig:F05}
Half-filled flat bands (i.e., F = $0.5$) when $U = 0.1 t$, 
corresponding to $\Delta_0 \approx 0.0253 t$ and $\mu \approx -2.00t$.
In (a), Goldstone contribution never merges with the quasiparticle 
continuum, reminiscent of typical BEC behavior. 
Its energy broadening is controlled by $\eta = 0.001t$. 
} 
\end{figure}
\begin{figure} [htb]
\includegraphics[width = 0.99\linewidth]{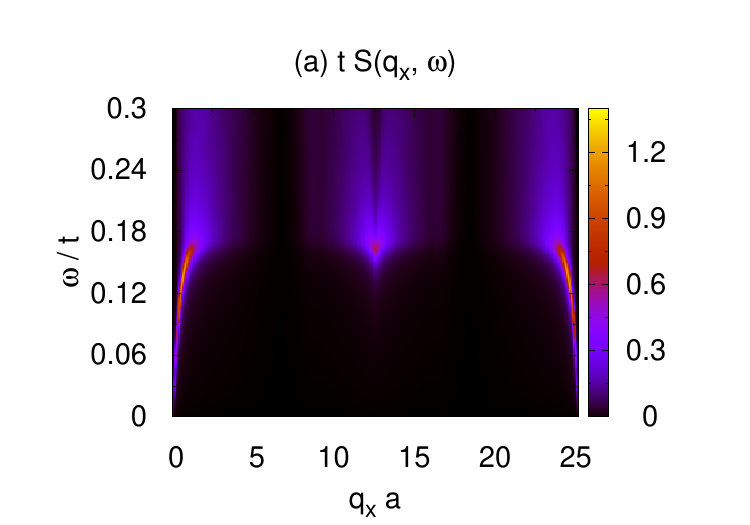}
\includegraphics[width = 0.99\linewidth]{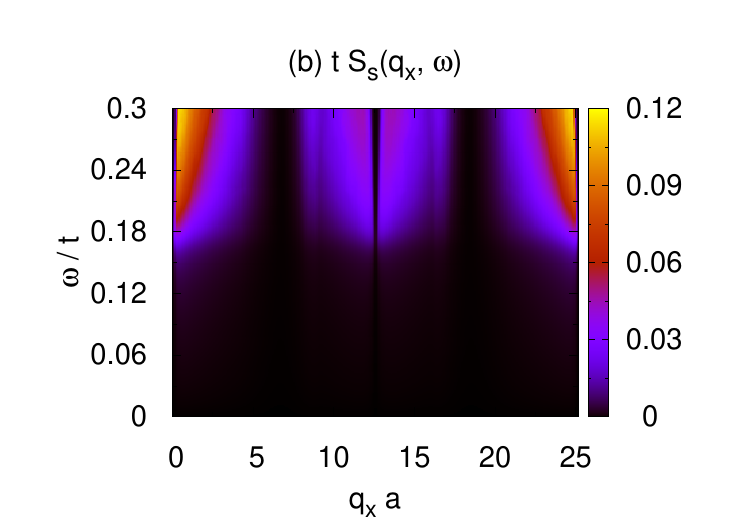}
\caption{\label{fig:F125}
Partially-filled dispersive bands (i.e., F = $1.25$) when $U = 2 t$,
corresponding to $\Delta_0 \approx 0.0831 t$ and $\mu \approx 0.247t$.
In (a), Goldstone contribution eventually merges with the quasiparticle 
continuum at high $\mathbf{q}$, reminiscent of typical BCS behavior.
Its energy broadening is controlled by $\eta = 0.01t$. 
} 
\end{figure}

First of all, when $U = 0$, the region $-2t < \mu < 6t$ lies within the 
dispersive bands and corresponds to particle fillings $1 < F < 2$, 
where $\mu = 2t$ is precisely at $F = 1.5$. On the other hand, while 
$\mu \to -2t$ from above corresponds to a half-filled lattice with $F \to 1$ 
from above, $\mu = -2t$ coincides with the degenerate flat bands when 
$0 < F < 1$. Focusing on the flat-band regime, we set $F = 0.5$
in Fig.~\ref{fig:F05}, corresponding to a half-filled flat bands, 
and present colored maps of the dynamic DSF
$
- \frac{1}{\pi} \mathrm{Im} \mathcal{X}(\mathbf{q}, \omega + i0^+)
$
and the dynamic SSF
$
- \frac{1}{\pi} \mathrm{Im} \mathcal{X}_{s}(\mathbf{q}, \omega + i0^+)
$
as functions of $\mathbf{q} = (q_x, 0, 0)$ and $\omega$ when $U = 0.1t$.
Here, the periodicity of the numerical results under 
$q_x \to q_x + 8\pi/a$ is a reflection of the periodicity of 
$h^{SS'}_{\mathbf{k} \sigma}$ under $k_x \to k_x + 8\pi/a$.
As the dynamic SSF is not coupled to the collective modes, i.e., 
see Eq.~(\ref{eqn:Xss}), it remains almost featureless at low frequencies 
until $\omega \sim \Theta$, marking the onset of the quasiparticle 
continuum in Fig.~\ref{fig:F05}(a), beyond which it exhibits significantly 
higher intensities.
On the other hand, the collective-mode contribution to Eq.~(\ref{eqn:Xnn}) 
is clearly visible as high-intensity peaks throughout momentum space, 
superimposed on a relatively low background intensity at both low and 
high frequencies. The fact that the collective mode never merges with 
the quasiparticle continuum is reminiscent of the BEC-like behavior in 
the usual BCS-BEC crossover of superfluid Fermi 
gases~\cite{zou16, hoinka17, biss22}. 
This result is also consistent with the Goldstone contribution in 
Eq.~(\ref{eqn:sfgm}), confirming that the Goldstone modes remain 
undamped even at high $\mathbf{q}$~\footnote{
Similar to the strong-coupling limit discussed in Sec.~\ref{sec:gc}, 
one can also show that in the dilute limit of the pyrochlore 
lattice, i.e., when $F \to 0$, the inequality
$
\bar{A} \bar{R} \ll \bar{B}^2
$ 
holds for any $U \ne 0$~\cite{iskin24a}. Furthermore, the superfluid-weight 
tensor again takes the form
$
\mathcal{D}_{ij} = 4 \rho_{Cp} (M_{Cp}^{-1})_{ij},
$
where both $\rho_{Cp}$ and $(M_{Cp}^{-1})_{ij}$ depend on $U$~\cite{iskin24a}.
As a result, Eq.~(\ref{eqn:sfgm}) also holds for any $U \ne 0$ in the 
dilute flat-band regime
}.
Note that, since we 
represent the Dirac delta function via a Lorentzian distribution
$
\delta(x) = \frac{1}{\pi} \lim_{\eta \to 0} \frac{\eta}{x^2 + \eta^2}
$
in our numerics, with $\eta = 0.001t$, the Goldstone modes possess some 
energy broadening. However, unlike the collective phase modes, the 
collective amplitude modes (i.e., the Higgs modes) do not leave any 
visible signature in the dynamic DSF, including the quasiparticle 
continuum region, where they are expected to decay into pairs of 
quasiparticle excitations. It is also worth mentioning that, 
similar to the in-phase Goldstone modes, the out-of-phase Leggett modes 
are expected to leave strong signatures in the dynamic DSF, at least 
in the strong-coupling limit where they remain undamped. However, a 
detailed analysis of these modes is beyond the scope of this paper.
In comparison to the flat-band results, we set $F = 1.25$ in  
Fig.~\ref{fig:F125}, corresponding to partially-filled dispersive bands
when $U = 2t$, and show that the collective mode is visible only at 
low $\mathbf{q}$ in the BCS regime, as it eventually merges with the  
quasiparticle continuum. This is reminiscent of the BCS-like behavior  
in the usual BCS-BEC crossover of superfluid Fermi  
gases~\cite{zou16, hoinka17, biss22}.

\section{Coulomb Interaction}
\label{sec:ci}

We now comment on the role of long-range Coulomb interactions among 
charged particles. In the weak-coupling limit of BCS theory, it is well
established that such interactions profoundly alter the collective excitation
spectrum. For instance, in three-dimensional systems, instead of 
supporting a gapless sound mode, 
the Coulomb interaction pushes the mode to high energies, 
resulting in a plasma oscillation that lies well above the superconducting 
gap. Consequently, these collective modes become largely irrelevant except 
for restoring gauge invariance. As interactions strengthen, however, the
superconducting gap increases while the plasma frequency decreases due 
to the growing effective mass $M_{Cp}$ of tightly bound Cooper pairs. 
In the extreme strong-coupling (BEC) limit, the system resembles a charged 
Bose gas, where the plasma frequency approaches zero as $M_{Cp}$ 
diverges. As discussed in Sec.~\ref{sec:ni}, this behavior also emerges 
in flat-band superconductors in the $U/t \to 0$ regime. Notably, in contrast 
to three dimensions, the plasma frequency in two dimensions scales as 
$\sqrt{|\mathbf{q}}|$ in the long-wavelength limit.

Incorporating Coulomb interactions into the Hubbard model is inherently
approximate, as the model treats $U$ as an effective attraction that already
accounts for screening and does not explicitly include the bare Coulomb 
potential. To address this, one typically employs the random-phase 
approximation (RPA)~\cite{belkhir94}, which yields the following
expression for the plasma frequency in the BEC regime of conventional 
superconductors:
\begin{align}
\label{eqn:plasmon}
\left(\omega_{\mathbf{q \to 0}}^P\right)^2 = V_{Cp}(\mathbf{q})
\rho_{Cp} \sum_{ij} \left(M_{Cp}^{-1}\right)_{ij} q_i q_j,
\end{align}
where 
$
V_{Cp}(\mathbf{q}) = 2(d-1)\pi e_{Cp}^2 / |\mathbf{q}|^{d-1}
$ 
is the Fourier transform of the Coulomb potential for Cooper pairs, 
$d = \{2,3\}$ is the dimensionality, and $e_{Cp} = 2e$ is their effective 
charge. The anisotropy in the plasmon dispersion, governed by the 
inverse effective-mass tensor $\mathbf{M}_{Cp}^{-1}$, is consistent with 
recent findings~\cite{ahn21}. In isotropic three-dimensional systems,
Eq.~(\ref{eqn:plasmon}) simplifies to the standard expression 
$4\pi e^2 \mathcal{D}$, also in agreement with the
literature~\cite{gabriele22}. Given our analysis in Sec.~\ref{sec:ni}, 
we expect Eq.~(\ref{eqn:plasmon}) to remain valid for flat-band 
superconductors, owing to their strongly interacting nature 
even in the $U/t \to 0$ regime.

\section{Conclusion}
\label{sec:conc}

In summary, we explored the interplay between quantum geometry and 
superconductivity in multiband BCS systems that exhibit time-reversal 
symmetry and uniform pairing. Using linear-response theory within the 
mean-field BCS-BEC crossover framework at $T = 0$, we analyzed the 
dynamic DSF and dynamic SSF of a multi-orbital Hubbard model. Our results 
rigorously satisfy the associated f-sum rules across various limits. 
Notably, in the strong-coupling regime, the behavior converges to that 
of a weakly-interacting Bose gas of Cooper pairs, with low-energy 
collective Goldstone modes emerging as Bogoliubov phonons. We further 
demonstrated that the quantum-geometric origin of these low-energy 
structure factors, and related observables such as the superfluid-weight 
tensor and the effective-mass tensor of Cooper pairs, can be traced 
back to the effective-mass theorem for Bloch bands. As an illustrative 
example, we investigate the pyrochlore-Hubbard model numerically, 
highlighting the profound impact of quantum geometry through collective
Goldstone modes on the superconducting properties of a complex 
flat-band lattice.
As an outlook, future studies could explore the contribution 
of collective Leggett modes to the dynamic DSF, as these out-of-phase 
modes are expected to play some role away from the strong-coupling 
limit toward the BCS-BEC crossover regime. For instance, in a multiband 
lattice with $N_b$ sublattices, we anticipate the emergence of $N_b-1$ 
distinct Leggett branches, all of which can be investigated numerically 
by introducing a sublattice-dependent order parameter for the 
fluctuations~\cite{iskin24a}.
Studying the effects of finite temperature is also crucial. However, 
even in a single-band lattice or continuum model, a significant 
theoretical challenge arises due to the presence of Landau singularities 
in the $(\mathbf{q}, \omega) \to (\mathbf{0}, 0)$ limit, which lead to 
the damping of even the low-energy collective modes~\cite{kurkjian19}.

\begin{acknowledgments}
The author acknowledges funding from US Air Force Office of Scientific 
Research (AFOSR) Grant No. FA8655-24-1-7391.
\end{acknowledgments}

\appendix

\section{Gaussian fluctuations}
\label{sec:gf}

In the case of multiband BCS superconductors with uniform pairing 
fluctuations, the effective Gaussian action can be 
written as~\cite{iskin24a, engelbrecht97, he16}
\begin{align}
\label{eqn:}
\mathcal{S}_\mathrm{GF} = \frac{1}{2} \sum_q 
\left( \Lambda_q^* \, \Lambda_{-q} \right)
\left( \begin{array}{cc}
\mathcal{M}_q^{11} & \mathcal{M}_q^{12} \\
\mathcal{M}_q^{21} & \mathcal{M}_q^{22}
\end{array} \right)
\left( \begin{array}{c}
\Lambda_q \\ \Lambda_{-q}^*
\end{array} \right),
\end{align}
where 
$
q \equiv (\mathbf{q}, iq_\ell)
$ 
is a combined index with $q_\ell = 2\ell \pi T$ the bosonic Matsubara
frequency, $\Lambda_q$ describes fluctuations of the order
parameter field around its saddle-point value $\Delta_0$, and the 
fluctuation matrix $\boldsymbol{\mathcal{M}}_q$ acts as an inverse 
propagator. Denoting 
$u^2 = (1+\xi/E)/2$, 
$u'^2 = (1+\xi'/E')/2$, 
$v^2 = (1-\xi/E)/2$ 
and $v'^2 = (1-\xi'/E')/2$, the matrix elements
$
\mathcal{M}_q^{12} = \mathcal{M}_q^{21} 
$
and
$
\mathcal{M}_q^{11} = \mathcal{M}_{-q}^{22}
$
can be written as
\begin{align}
\mathcal{M}_q^{11} &= \frac{N_S}{U} + \frac{1}{N_c} 
\sum_{nm \mathbf{k}} 
\left( \frac{u^2 u'^2}{iq_\ell-E-E'} 
- \frac{v^2 v'^2}{iq_\ell+E+E'} \right)Z, \\
\mathcal{M}_q^{12} &= \frac{1}{N_c} \sum_{nm \mathbf{k}} 
\left( \frac{u v u 'v'}{iq_\ell+E+E'} 
- \frac{u v u'v'}{iq_\ell-E-E'} \right)Z,
\end{align}
at $T = 0$. Next we introduce the transformation
$
\Lambda_q = \lambda_{1,q} + i \lambda_{2,q},
$
and associate $\lambda_{1,q} = \lambda_{1,-q}^*$ and 
$\lambda_{2,q} = \lambda_{2,-q}^*$ with the amplitude and phase 
degrees of freedom, respectively. This leads to the following 
expressions for the effective couplings in the Gaussian action:
$
I_q^{11} = \mathcal{M}_q^{11} + \mathcal{M}_q^{22} 
+ \mathcal{M}_q^{12} + \mathcal{M}_q^{21}
$
for the amplitude-amplitude,
$
I_q^{22}  = \mathcal{M}_q^{11} + \mathcal{M}_q^{22} 
- \mathcal{M}_q^{12} - \mathcal{M}_q^{21}
$
for the phase-phase, and
$
q_\ell I_q^{12}  = i(\mathcal{M}_q^{11} - \mathcal{M}_q^{22} 
- \mathcal{M}_q^{12} + \mathcal{M}_q^{21})
$
for the amplitude-phase. With these, the effective Gaussian 
action becomes
\begin{align}
\label{eqn:}
\mathcal{S}_\mathrm{GF} = \frac{1}{2} \sum_q \left( \lambda_{1,-q} \, \lambda_{2,-q} \right)
\left( \begin{array}{cc}
I_q^{11} & q_\ell I_q^{12}  \\
-q_\ell I_q^{12}  & I_q^{22} 
\end{array} \right)
\left( \begin{array}{c}
\lambda_{1,q} \\ \lambda_{2,q}
\end{array} \right).
\end{align}
In Sec.~\ref{sec:gm}, $I_{\mathbf{q}\omega}^{ij}$ denotes $I_q^{ij}$ after 
analytic continuation $iq_\ell \to \omega$.

\section{Ward identities}
\label{sec:wi}

The Ward identities for the density response functions given in 
Ref.~\cite{guo13} can be easily generalized to the case of multiband 
BCS superconductors. At $T = 0$, their Eq.~(42) is analogous to the 
following expressions 
\begin{widetext}
\begin{align}
\frac{1}{N_c} \sum_{nm \mathbf{k}} 
\frac{E+E'}{EE'} \frac{(\xi + \xi')Z}{\omega^2 - (E+E')^2}
- \frac{2}{N_c} \sum_{nm \mathbf{k}} 
\bigg(\frac{\xi}{E} + \frac{\xi'}{E'}\bigg)  
\frac{Z}{\omega^2 - (E+E')^2}
= \frac{1}{N_c} \sum_{nm \mathbf{k}} 
\frac{E-E'}{EE'} \frac{(\xi - \xi')Z}{\omega^2 - (E+E')^2},
\\
\frac{1}{N_c} \sum_{nm \mathbf{k}} 
\frac{E + E'}{EE'} \frac{\omega^2 Z}{\omega^2 - (E+E')^2}
- 2\bigg[
\frac{2N_b}{U} + \frac{1}{N_c}  \sum_{nm \mathbf{k}}
\frac{E + E'}{EE'} \frac{EE' + \xi\xi' + \Delta_0^2}{\omega^2 - (E+E')^2} Z
\bigg]
= \frac{1}{N_c} \sum_{nm \mathbf{k}} 
\frac{E + E'}{EE'} \frac{(\xi-\xi')^2 Z}{\omega^2 - (E+E')^2},
\\
\frac{1}{N_c} \sum_{nm \mathbf{k}} 
\frac{E + E'}{EE'} \frac{EE' - \xi\xi' + \Delta_0^2}{\omega^2 - (E+E')^2} Z
- \frac{1}{N_c} \sum_{nm \mathbf{k}} 
\frac{E + E'}{EE'} \frac{2\Delta_0^2 Z}{\omega^2 - (E+E')^2}
= \frac{1}{N_c} \sum_{nm \mathbf{k}} 
\bigg(\frac{\xi}{E} - \frac{\xi'}{E'}\bigg)  
\frac{(\xi - \xi')Z}{\omega^2 - (E+E')^2}.
\end{align}
\end{widetext}
These identities can be verified through algebraic manipulations, e.g., 
the second identity can be shown via the substitution
$
4N/U = \sum_{nm \mathbf{k}} (Z/E + Z/E'),
$
which follows from the order-parameter Eq.~(\ref{eqn:op}).

\section{Effective-mass tensors}
\label{sec:emt}

Note that the inverse effective-mass tensor $\mathbf{M}_{n\mathbf{k}}^{-1}$ 
for the $n$th Bloch band~\cite{iskin19b} and the inverse effective-mass tensor 
$\mathbf{M}_{2b}^{-1}$ for the lowest-lying two-body bound 
states~\cite{iskin22} can be written compactly as
\begin{align}
(M_{n\mathbf{k}}^{-1})_{ij} &= \sum_{m} (f_{nm\mathbf{k}}^{-1})_{ij}, \\
(M_{2b}^{-1})_{ij} &= 
\frac{\sum_{nm\mathbf{k}} \frac{(f_{nm\mathbf{k}}^{-1})_{ij} - 
\ddot{\varepsilon}_{n\mathbf{k}}^{ij} \delta_{nm}/2}
{(2\varepsilon_{n \mathbf{k}} - E_b)
(\varepsilon_{n \mathbf{k}} + \varepsilon_{m \mathbf{k}} - E_b)}}
{\sum_{n\mathbf{k}} \frac{1}{(2\varepsilon_{n \mathbf{k}} - E_b)^2}},
\label{eqn:mt2b}
\end{align}
where $E_b < 0$, with $|E_b|$ denoting the binding energy of the pair, and
\begin{align}
(f_{nm\mathbf{k}}^{-1})_{ij} = \ddot{\varepsilon}_{n\mathbf{k}}^{ij} \delta_{nm}
- (\varepsilon_{n \mathbf{k}} - \varepsilon_{m \mathbf{k}})
g_{ij}^{nm\mathbf{k}}
\end{align}
can be interpreted as the band-resolved inverse effective-mass tensor
for the $n$th Bloch band. Then, it is again reassuring to verify that 
Eq.~(\ref{eqn:mt2b}) is in perfect agreement with Eq.~(\ref{eqn:Mcp}) 
in the $U \gg t$ limit, where $|E_b| \to U \gg \max\{2\varepsilon_{n \mathbf{k}}\}$. 
To show this, we make use of the following identities:
$
\sum_{n \mathbf{k}}  \ddot{\varepsilon}_{n\mathbf{k}}^{ij} = 0,
$
$
g_{ij}^{nm\mathbf{k}} = g_{ji}^{mn\mathbf{k}},
$
$
g_{ij}^{n\mathbf{k}} = g_{ji}^{n\mathbf{k}}
$
and
$
\sum_{n, m\ne n} \varepsilon_{m\mathbf{k}} g^{nm\mathbf{k}}_{ij} = 
\sum_{n} \varepsilon_{n\mathbf{k}} g^{n\mathbf{k}}_{ij}.
$
This correspondence further illustrates the quantum-geometric origin 
of the pair mass. Notably, a similar geometric contribution has also been
reported for the optical mass~\cite{yu24}.

\bibliography{refs}

\end{document}